# An Enhanced Framework with Advanced Study to Incorporate the Searching of E-Commerce Products Using Modernization of Database Queries

Mohd Muntjir
College of Computers and Information Technology
Taif University
Taif, Saudi Arabia

Ahmad Tasnim Siddiqui
College of Computers and Information Technology
Taif University
Taif, Saudi Arabia

*Abstract*—This study aims to inspect and evaluate the integration of database queries and their use in e-commerce product searches. It has been observed that e-commerce is one of the most prominent trends, which have been emerged in the business world, for the past decade. E-commerce has gained tremendous popularity, as it offers higher flexibility, cost efficiency, effectiveness, and convenience, to both, consumers and businesses. Large number of retailing companies has adopted this technology, in order to expand their operations, across of the globe; hence they needs to have highly responsive and integrated databases. In this regard, the approach of database queries is found to be the most appropriate and adequate techniques, as it simplifies the searches of e-commerce products.

## I. INTRODUCTION

The purpose of this paper is to present the illustration of database queries as well as their use in searching e-commerce products. It has been assessed that the concept of e-commerce has gained tremendous popularity with the emergence of innovative and advanced technological tools. According to Li and Karahanna (2012), e-commerce or electronic commerce can be understood as the trading of services or products, by using computer networks, usually internet. It has been assessed that the functions or operations of e-commerce draws on different technologies, including automated data collection systems, inventory management systems, EDI (electronic data interchange), electronic funds transfer ,online transaction processing, supply chain management ,internet marketing, and mobile commerce. In this regard, Liu, et.al, (2010) has declared that contemporary e-commerce utilizes the World Wide Web, in order to conduct different transactions and retailing activities.

Recently researchers have developed various different techniques, in order to access and purchase e-commerce products and services. In this account, Endrullis, et.al, (2012) has asserted that integration and utilization of database queries may play a commendable role in searching e-commerce products, more conveniently. Database queries can be understood as one of the most advanced databases, which are based on the relational model, which was established by Codd. It is significant to notice that query form is found to be one of the most efficient and integrated user interfaces, which are being widely used for querying databases (VanderMeer, et.al, 2012). It has been inspected that database queries play an inevitable role in providing feasible and quick access to the required information or products.

It is due to the fact that in this paradigm, a database table usually represents a mathematical relationship amid set of different products or items, having similar attributes or characteristics. In other words, the entire framework of database queries is developed in such a manner, which assists its use to access ample amount of relevant and creditable information from complex and large databases. Thereby, it can be avowed that the integration and use of database queries may play an indispensable role in e-commerce product searches (Vander Meer, et.al, 2012). The proceeding paper will help in understanding the core concept of e-commerce as well as database queries. Mainly this study will focus on the integration of database queries in the process of searching e-commerce products.

## II. AIM AND OBJECTIVES

This study aims to investigate the integration and use of database queries in e-commerce product searches. Proceeding mentioned objectives would be fulfilled, in order to accomplish this research aim.

- To analyze the notion of e-commerce;
- To understand the basic concept of database queries;
- To examine different modules of database queries;
- To assess different techniques of integrating and utilizing database queries in order to search e-commerce products.

## III. SIGNIFICANCE OF RESEARCH

According to Das-Sarma, et.al, (2014), technological developments have brought considerable changes in the lives of people, in terms of performing daily routine tasks. Most prominently, these technological developments have influenced business and retailing sector. In this regard, e-commerce can be considered as one of the most prominent advancements, which have been occurred in retailing industry, due to technological advancements. In this regard, Liu, et.al, (2010) has claimed that e-commerce has played a major role in transforming the actual faced of retailing industry, as this innovative approach allows the users to perform their transactions and purchasing activities by using different online and digital tools, regardless of their geographical locations (Telang, et.al, 2012).

It has been documented in the studies of Endrullis, et.al, (2012) in recent times; e-commerce has been developed at fast paced, across the world. E-commerce can be understood as the process of purchasing or selling services and goods, by using electronic tools. It is important to bring into the notice that e-commerce transactions can be conducted amid private and public organizations, governments, individuals, households, and businesses. It has been established that there are various different types of e-commerce transactions that takes place online ranging from sale of books, shoes, cloths to different services, like making hotel bookings or airline tickets.

During e-commerce activities, users have to face difficulties, in terms of searching their desired products and services. In this scenario, e-businesses have to implement and integrate high-tech systems, in order being ease and convenience for e-commerce users, in terms of searching their required products and services. It has been claimed by Li and Karahanna (2012) that database queries are found to be one of the integrated and valuable methods, which assists the users in accessing required products, in considerably massive and sophisticated databases.

Basically, the technique of database queries helps in optimizing the products through keywords; hence results in immediate and quick product searches. Thereby, it can be affirmed that the integration of database queries in e-commerce product searches is one of the greatest initiatives towards integrated and sustainable retailing activities. However, this technique also possesses various disadvantages, most prominently language gap. It has been established that the language gap is usually occurred in the keywords, which are used for search queries, and database's product specifications (Telang, et.al, 2012).

## IV. LITERATIRE REVIEW

### A. E-Commerce-The Concept

Li and Karahanna (2012) have claimed that e-commerce includes several activities, including exchanging, selling, and buying of services, products, and information, through computer networks, mainly internet. It is important to notice that the term "commerce", is usually referred to the transactions, which are conducted amid business partners. When the concept of e-commerce was evaluated on the pre-discussed definition of commerce, it was recognized that e-commerce is the same as electronic business or e-business, but this approach is not true. It is due to the fact that e-commerce does not only deal with selling or purchasing of goods, but also with collaborating with servicing customers, business partners, and performing electronic transactions within an association.

Accordingly, search marketing success for e-commerce sites is predicated less on innovation than possessing an expert understanding of factors that impact search marketing efforts.

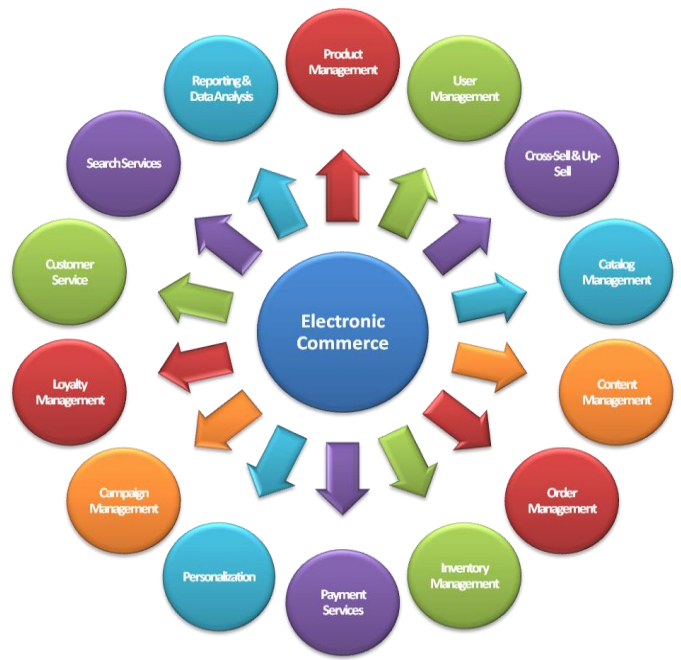

Fig. 1. Classifications of E-Commerce

It has been documented in the researches and studies, which were conducted by Liu, et.al, (2010) that e-business is all about reaching new customers, enhanced productivity, globalization, speed, time cycle, and sharing knowledge across institutions, in order to gain competitive advantages. According to Endrullis, et.al, (2012), e-commerce offers wide range of benefits to the organizations, customers, as well as to the entire society. When e-commerce benefits, for organizations, were analyzed it was revealed that e-commerce enables the businesses and companies to expand their market to international and national levels, while saving their cost and time.

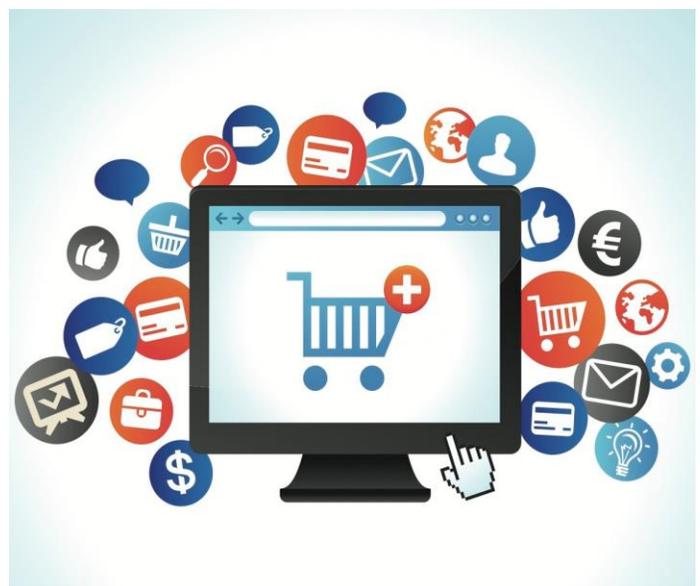

Fig. 2. E-Commerce

In this account, Leszczynski and Stencel (2011) have affirmed that e-commerce plays an appreciable role in quickly locating more customers, suitable business partners, as well as valuable suppliers for business. In addition to this, e-commerce also allows the companies to procure services and material from other companies, in a cost effective and efficient manner. E-commerce also enables considerably specialized niche market. In customer's perspective, e-commerce provides less expensive services and products by enabling the customers to perform quick online comparisons.

More so, e-commerce also helps the consumers in getting customized products, ranging from computer systems to super luxury cars, at highly competitive prices. According to Leszczynski and Stencel (2011), e-commerce is also found to be one of the most beneficial approaches for the entire society. It is due to the fact that e-commerce allows people, living in remote areas, to get their desired products, while saving travelling cost. Furthermore, e-commerce also assists people to receive well-timed healthcare services, education facilities, etc., irrespective of their geographical locations.

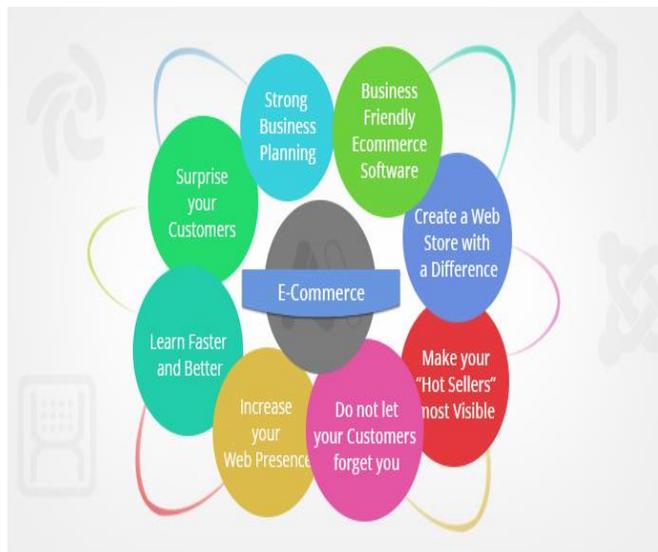

Fig. 3. E-Commerce Framework

*B. Database Queries*

It has been established from the analysis of research and studies, which were carried out by Liu, et.al, (2010) that database queries can be considered as one of the most effective and integrated approaches, which allow its users to attain their desired information or data from substantially gigantic databases. It has been identified that the technique of database queries plays an inevitable and incredible role in enabling the users to point out their desired information, without accessing or making efforts to search for the entire table. Endrullis, et.al, (2012) has supported this idea by claiming that the approach of database queries allows its users to amalgamate wide range of tables. This phenomenon can be easily understood by considering an example, which is, if a user of the database is dealing with two different tables, including consumers and invoices, they can easily use this technique, i.e., database queries.

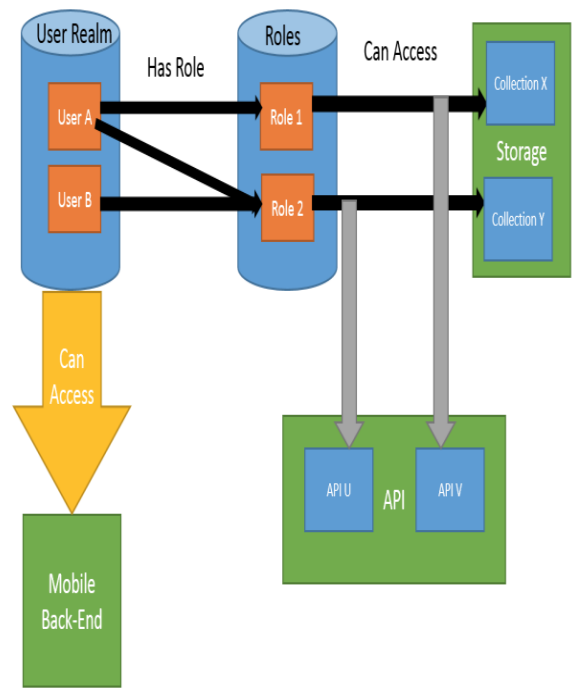

Fig. 4. A Framework of a System To Fetch the Data

It is due to the fact that this database queries inevitably helps in amalgamating the contents or information, which is being stored in two different tables. Afterwards, the names and complete data of the customers can be easily attained, by executing this query by the user. It is important to notice that the final results are according to the invoices of the consumers. Leszczynski and Stencel (2011) have presented an idea according to which the technique of database queries is not capable enough to store the data. Lu, et.al, (2013) has supported this approach by asserting that database queries can only identify the stored data, instead of storing it.

In accordance with the views of Telang, et.al, (2012) database query possesses wide range of benefits, which may play a major role in impacting the operations of e-commerce. In this regard, one of the most prominent advantages of database queries is that it merges or amalgamates wide range of data or information, which is being stored in different databases. In other words, database query combines required and valuable information from several different sources of data. In addition to this, database query also enable its users to choose their required fields from different sources, while identifying them, as per their needs. More so, this innovative and integrated framework also helps in specifying the records, which are similar to the criteria, which was predefined by the users (Li, et.al, 2013).

It has been documented in the researches of Endrullis, et.al, (2012), apart from various advantages; database queries also comprise different disadvantages. One of the most prominent disadvantages of database queries may include the language gap amid the requests (made by the users) and technical terms, which are used during the development of a database. It has been observed that researchers are intending to cope with this issue, as this feature considerably affects the reliability of database queries (VanderMeer, et.al, 2012).

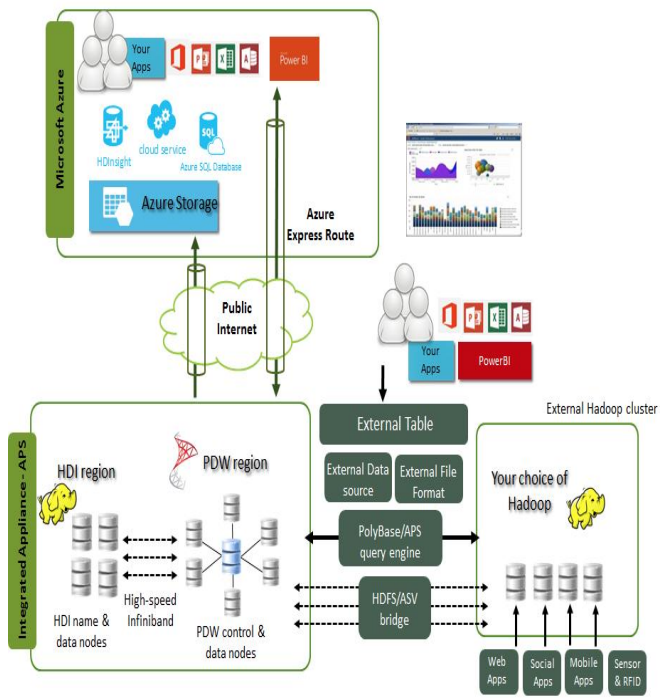

Fig. 5. An Architecture of Database To Fetch The Queries

### C. Modules of Database Queries

It has been claimed by Leszczynski and Stencel (2011) that various modules have been developed by the researchers, in order to ensure the integrity and credibility of the database queries' operations. Some of the most prominent modules are briefly described in the proceeding paper.There are given two modules for Database Queries.

*a) Query Breaking Module:* According to Das-Sarma, et.al, (2014), query breaking module plays a crucial role in ensuring integrated operations of database query. It has been analyzed that massive queries to database often results in the malfunctioning of the database; hence results in stoppage of database or other unfavorable consequences. Therefore, it isnecessary to break the chosen queries into large number of small queries. It has been observed that the activity plays an inevitable and incredible role in reducing potential risks, which are usually occurred due to increased complexities. Li and Karahanna (2012) have asserted that the entire process can be easily conducted by the help of query breaking module. However, this module also possesses some issues, in terms of identifying adequate LECO (local e-catalog ontologies).

*b) Query Reasoning and Expanding Module:*

Query reasoning and expanding module is another most effective feature, which resides in database queries. It has been observed that this module is entirely different from traditional query system, as the primary features of the semantic query include the activity of expanding the query reason, during the process of querying. This function helps the users in identifying and attaining their desired information and statistics, in anappropriate and effectivebehavior (Wang, et.al, 2012).

### V. INTEGRATING AND UTILIZING DATABASE QUERIES IN E-COMMERCE PRODUCT SEARCHES

In the current era, businesses are continually starving to expand their operations, across the globe, while controlling their operational cost.

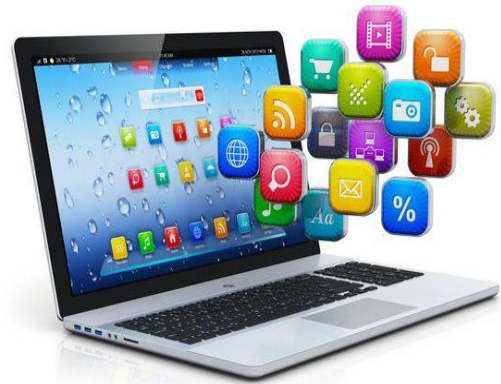

Fig. 6. Demonstration of E-Commerce Product Search

In this regard, the emergence of e-commerce has commendably benefited and supported the businesses, in terms of expanding their operations, on local and international levels (Wang, et.al, 2012). It is a fact that e-commerce has played a vital role in facilitating the businesses as well as consumers in terms of performing their transactions and retailing activities. On the other hand, continually increasing use of e-commerce has also affected the activities of product searching. It is due to the fact that massive number of requests (queries), made by the consumers often affect the performance and functionality of the databases; hence influencing the integrity of e-commerce activities. In this account, Das-Sarma, et.al, (2014) has avowed that continuously increasing demands of e-commerce have pressurized the database developers to formulate such databases, which are capable enough to directly find out the query of the webpage content. In this scenario, the approach of database queries can be characterized amid one of the most appropriate solutions.

Recent studies, which are accomplished by Li, et.al, (2013) have revealed the fact that e-commerce has made it feasible for people to shop their desired products and services, in spite of living in remote areas. Emerging trends of e-commerce have also played an indispensable role in supporting mid as well as small sized companies to increase their revenues or profits through e-commerce. It is because; this tool (e-commerce) helps them in expanding their target market, without investing large capitals (Lu, et.al, 2013).

Therefore, it can be affirmed that e-commerce has opened new doors of feasibility, convenience, cost efficiency, and time efficiency, both for companies and consumers. While considering these trends, it can be stated that the integration of database queries in e-commerce activities may result in commendably profitable outcomes. It is due to the fact that database queries may considerably reduce certain ambiguities and indistinctness, while carrying out e-commerce product searches (Wang, et.al, 2012).

It has been asserted by Das-Sarma, et.al, (2014) that database queries may also facilitate various other areas of life, including ERP systems, CRM systems, personalized marketing applications, e-commerce apps, as well as custom-built apps. It is significant to notice that database queries encapsulate the traits of relational databases, which plays a noticeable role in improving the overall performance of the product search (Lu, et.al, 2013).

When database queries are incorporated with the e-commerce product searches, it usually results in higher atomicity, consistency, integrity, durability, accuracy, and higher efficacy, in terms of transactions. All of these features are considered as most important elements for the environment of e-commerce product searches. Therefore, it can be stated that integration of database queries with e-commerce product searches may commendablysupport the companies, which have adopted e-commerce approach. It has been observed that the paradigm of database queries inevitably improves the scalability and reliability of e-commerce product search operations and aidsthe consumers to find out their preferred products and services, in an accurate and efficient manner (Li, et.al, 2013).

Endrullis, et.al, (2012) has claimed that consumers have to submit their vouchers or payment information, while purchasing their desired products and services, from different websites (e-commerce websites). In this regard, the consumers have to choose their required items, while assessing the specialties (cost, quality, etc.) of the products. On the other hand, companies have to identify the requirements, demands, and current trends of the market. In this scenario, integrated and systematically developed database query systems may substantially help them in developing their products as well as updating their e-commerce websites (Lu, et.al, 2013).

## VI. Conclusion

From above discussion, it can be concluded that integration of database queries in e-commerce product searches is one of the greatest initiatives towards integrated, smooth, and systematic e-commerce activities. It is due to the fact that the technique of database queries plays an indispensable role in eliminating the probability of different risks, which are often occurred in different databases. Highly advanced and innovative modules of database queries play a vital role in controlling excessive traffic of queries, by breaking large queries into small ones; hence results in more efficient and well-timed searches of e-commerce products. The proceeding paper has briefly discussed the concept of e-commerce and database queries. More so, the paper has also encapsulated the analysis of different aspects, which are associated with the integration of database queries and their use in e-commerce product searches.

## Acknowledgment

We would like to gratefully and sincerely thank to The Dean of our College, and Chairman of our Departments for their guidance, understanding, patience, and most importantly, their friendly nature during this research paper writing at College of computers and Information Technology, Taif University, Saudi Arabia. We would also like to thank all of the members of the research group and friends who have always supported us for giving the opportunity to write this research paper.